\documentclass[aps,prd,showpacs,floats]{revtex4}
\usepackage{epsfig}
\usepackage{dcolumn}
\usepackage{bm}
\usepackage{amsmath}
\usepackage{amsfonts}
\usepackage{amssymb}
\usepackage{graphicx}

\newcommand{\bea}{\begin{eqnarray}}
\newcommand{\eea}{\end{eqnarray}}
\newcommand{\bwt}{\begin{widetext}}
\newcommand{\ewt}{\end{widetext}}
\def\be{\begin{equation}}
\def\ee{\end{equation}}
\begin{document}

\title{Confront \textit{holographic} QCD with Regge trajectories of vectors
and axial-vectors }
\author{Song He$^{1}$}
\author{Mei Huang$^{1,2}$}
\author{Qi-Shu Yan$^{3,4}$}
\author{Yi Yang$^{5,6}$}
\affiliation{$^{1}$ Institute of High Energy Physics, Chinese Academy of Sciences,
Beijing, China }
\affiliation{$^{2}$ Theoretical Physics Center for Science Facilities, Chinese Academy of
Sciences, Beijing, China}
\affiliation{$^{3}$ Department of Physics, University of Toronto, Toronto, Canada}
\affiliation{$^{4}$ Department of Physics, National Tsing Hua University, Hsinchu, Taiwan}
\affiliation{$^{5}$ Department of Electrophysics, National Chiao-Tung University,
Hsinchu, Taiwan}
\affiliation{$^{6}$ Physics Division, National Center for Theoretical Sciences, Hsinchu,
Taiwan}
\date{\today }

\begin{abstract}
We derive the general 5-dimension metric structure of the $Dp-Dq$ system in
type II superstring theory, and demonstrate the physical meaning of the
parameters characterizing the 5-dimension metric structure of the \textit{%
holographic} QCD model by relating them to the parameters describing Regge
trajectories. By matching the spectra of vector mesons $\rho_1$ with
deformed $Dp-Dq$ soft-wall model, we find that the spectra of vector mesons $%
\rho_1$ can be described very well in the soft-wall $D3-Dq$ model, i.e, $%
AdS_5$ soft-wall model. We then investigate how well the $AdS_5$ soft-wall
model can describe the Regge trajectory of axial-vector mesons $a_1$. We
find that the constant component of the 5-dimension mass square of
axial-vector mesons plays an efficient role to realize the chiral symmetry
breaking in the vacuum, and a small negative $z^4$ correction in the 5-dimension 
mass square is helpful to realize the chiral symmetry restoration in high
excitation states.
\end{abstract}

\pacs{11.25.Tq, 11.10.Kk, 11.25.Wx, 12.38.Cy}
\maketitle


\section{Introduction}

Quantum Chromodynamics (QCD) has been accepted as the basic theory of
describing strong interaction for more than 30 years. However, it is still a
challenge to solve QCD in non-perturbative region where gauge interaction is
strong. In the early 1970's, string theory was proposed to describe strong
interacting particles \cite{string-1970}. Recently, the discovery of the
gravity/gauge duality \cite{dual} has revived the hope to understand QCD in
strongly coupled region using string theory. The gravity/gauge, or anti-de
Sitter/conformal field theory (AdS/CFT) correspondence provides a
revolutionary method to tackle the problem of strongly coupled gauge
theories. For a review of AdS/CFT, see \cite{AdS/CFT}. The string
description of realistic QCD has not been successfully formulated yet. Many
efforts are invested in searching for such a realistic description by using
the "top-down" approach, \textit{i.e.} by deriving holographic QCD from
string theory \cite{nonAdS-QCD}, as well as by using the "bottom-up"
approach, \textit{i.e.} by examining possible \textit{holographic} QCD
models from experimental data \cite%
{Son:2003et,TB:05,DaRold2005,EKSS2005,Ghoroku:2005vt}.

It is an essential and crucial point for the realistic \textit{holographic}
QCD model to reproduce Regge behavior. Regge behavior is a well-known
feature of QCD \cite{Regge}, and it was the commanding evidence for
suggesting the string-like structure of hadrons. A general empirical
expression for Regge trajectories can be cast as 
\begin{align}
M^{2}_{n,S} & = a_{n} \, n \, + \, a_{S} \, S + b \,,  \label{ab}
\end{align}
where $n$ and $S$ are the quantum number of high radial and spin
excitations, respectively. The slope $a_{n}$ and $a_S$ have dimension $%
\mathrm{GeV}^2$, and describe the mass square increase rate in radial
excitations and spin excitations, respectively. In principle, $a_{n}$ is not
necessarily the same as $a_{S}$, though $a_{n}=a_{S}$ can be taken as a good
approximation by fitting experimental data \cite{Anisovich:2000kx}. The
parameter $b$ is the ground state mass square, and it is channel-dependent.

Currently, one of the most successful models derived from string theory is
the Sakai-Sugimoto model \cite{SS}, which can describe spontaneously chiral
symmetry breaking naturally, but fails to generate the linear Regge
behavior. On the other hand, in the "bottom-up" approach, many efforts have
been paid to generate the linear Regge behavior for meson spectra \cite%
{KKSS2006,and06,kru05} and for baryon spectra \cite{Forkel}. In Ref. \cite%
{KKSS2006}, Karch, Katz, Son and Stephanov (KKSS) found that a $z^{2}$
dilaton field correction in the $AdS_5$ background leads to the linear
behavior $M_{n}^{2} \propto n$.

In this paper, we carefully study realizing Regge trajectories of vector and
axial-vector mesons in \textit{holographic} QCD model. We take the data of
the radial and spin excitations of $\rho$ and $a$ from PDG2007 \cite{pdg2007}%
, which are listed in Table \ref{tablerhoa}. 
\begin{table}[th]
\begin{center}
\begin{tabular}{|c|c|c|c|c|c|c|}
\hline
n & $1$ & $2$ & $3$ & $4$ & $5$ & $6$ \\ \hline
$1^{--}_{\rho^{n}}$ & $0.770$ & $1.450$ & $1.700$ & $1.900$ & $2.150$ & $%
2.270$ \\ \hline
$3^{--}_{\rho^{n}}$ & $1.690$ & $1.990$ & $2.250$ & $-$ & $-$ & $-$ \\ \hline
$1^{++}_{a_{1}^{n}}$ & $1.260$ & $1.640$ & $1.930$ & $2.095$ & $2.270$ & $%
2.340$ \\ \hline
$3^{++}_{a_{3}^{n}}$ & $1.870$ & $2.070$ & $2.310$ & $-$ & $-$ & $-$ \\ 
\hline
\end{tabular}%
\end{center}
\caption{Vector and axial-vector meson spectra (in GeV).}
\label{tablerhoa}
\end{table}
To describe Regge trajectories for both ($\rho_{1}$, $\rho_{3}$) and ($a_{1}$%
, $a_{3}$), we use the general formula Eq. (\ref{ab}). From the experimental
data, the parameters of Regge trajectories can be determined by using the
standard $\chi^{2}$ fit. The parameters for $(\rho_1,\rho_3)$ mesons and
their correlations read 
\begin{equation}
\begin{array}{rl}
a_{n}^{\rho}= \!\!\! & +0.91\pm0.23 \\ 
a_{S}^{\rho}= \!\!\! & +1.08\pm0.39 \\ 
b^{\rho}= \!\!\! & -1.09\pm1.18%
\end{array}
\begin{pmatrix}
1 &  & \cr -0.82 & 1 & \cr +0.43 & -0.81 & 1%
\end{pmatrix}
.  \label{eq:rhoregge}
\end{equation}
The parameters for $(a_1,a_3)$ mesons and their correlations read 
\begin{equation}
\begin{array}{rl}
a_{n}^{a} = \!\!\! & +0.81\pm0.22 \\ 
a_{S}^{a} = \!\!\! & +0.88\pm0.39 \\ 
b^{a} = \!\!\! & +0.13\pm1.17%
\end{array}
\begin{pmatrix}
1 &  & \cr -0.82 & 1 & \cr +0.43 & -0.81 & 1%
\end{pmatrix}
.  \label{eq:aregge}
\end{equation}

From the Regge trajectories of $\rho$ and $a$, one can read the information
of chiral symmetry breaking in the vacuum and asymptotic chiral symmetry
restoration in highly excited meson states:

\textit{Chiral symmetry breaking:} It is known that chiral symmetry is
spontaneously broken in the vacuum, thus the observed chiral partners are
not degenerate. From the Regge trajectories of the chiral partners $\rho$
and $a$, the chiral symmetry breaking in the vacuum is reflected by the
difference of the ground state square-masses $b^{\rho}$ and $b^{a}$. The
difference is as large as $1~ \mathrm{GeV}^2$, \textit{i.e.} $%
b^{a}-b^{\rho}\simeq M_{a_1}^2-M_{\rho_1}^2 \simeq 1~\mathrm{GeV}^2$.

\textit{Asymptotic chiral symmetry restoration in highly excited states:} It
is noticed that the chiral partner becomes more and more degenerate in high
excitations \cite{chiralbr}. For example, for radial excitations of $\rho_1$
and $a_1$, the mass square difference of chiral partners $%
M_{a_1}^2-M_{\rho_1}^2=0.9947~\mathrm{GeV}^2$ at $n=1$, and this difference
decreases to $0.3227~ \mathrm{GeV}^2$ at $n=6$; For the radial excitations
of $\rho_3$ and $a_3$, the mass square difference of chiral partners $%
M_{a_3}^2-M_{\rho_3}^2=0.6408~ \mathrm{GeV}^2$ at $n=1$, and it decreases to 
$0.2736~ \mathrm{GeV}^2$ at $n=3$.

We will only focus on the realization of Regge trajectories of $\rho_1$ and $%
a_1$ in this paper. The paper is organized as following: After the
introduction, we derive the general 5-dimension metric structure of the $%
Dp-Dq$ system in type II superstring theory in Sec. \ref{sec-DpDq}. By
matching the spectra of vector mesons $\rho_1$ with deformed $Dp-Dq$
soft-wall model, in Sec. \ref{sec-rho-metric} we determine the 5-dimension
metric that can describe the spectra of vector mesons $\rho_1$. We then
investigate how to describe the spectra of axial-vector mesons $a_1$ in the
same background of $\rho_1$ in Sec. \ref{sec-a-metric}. At the end we give
summary in Sec.\ref{sec-summary}.

\section{5-dimension metric structure of $D_p-D_q$ soft-wall model}

\label{sec-DpDq}

\subsection{Metric structure of $D_p-D_q$ system}

In order to investigate the possible dual string theory for describing Regge
behavior, we introduce the following $Dp-Dq$ branes system in type II
superstring theory. In the $Dp-Dq$ system, the $N_{c}$\ background $Dp$%
-brane describe the effects of pure QCD theory, while the $N_{f}$ probe\ $Dq$%
-brane is to accommodate the fundamental flavors which has been introduced
by Karch and Katz \cite{Karch:2002sh}. Such a practice is well motivated
from string theory side. For example, in Sakai-Sugimoto model, $p=4$ and $%
q=8 $. Low energy hadronic excitations are fields on probe branes which in
the background determined by the background branes.

First, we consider $N_{c}$\ background $Dp$-branes in type II superstring
theory. The near horizon solution in 10-dimension space-time is \cite%
{Dp-brane}%
\begin{equation}
ds^{2}=h^{-\frac{1}{2}}\eta _{\mu \nu }dx^{\mu }dx^{\nu }+h^{\frac{1}{2}%
}\left( du^{2}+u^{2}d\Omega _{8-p}^{2}\right) ,  \label{metric}
\end{equation}%
where\ $\mu ,\nu =0,\cdots ,p$, the warp factor $h\left( u\right) =\left(
R/u\right) ^{7-p}$ and $R$ is a constant%
\begin{equation}
R=\left[ 2^{5-p}\pi ^{\left( 5-p\right) /2}\Gamma \left( \frac{7-p}{2}%
\right) g_{s}N_{c}l_{s}^{7-p}\right] ^{\frac{1}{7-p}}.
\end{equation}%
The dilaton field in this background has the form of 
\begin{equation}
e^{\Phi }=g_{s}~h\left( u\right) ^{\frac{\left( p-3\right) }{4}}.
\label{dilaton}
\end{equation}%
The effective coupling of the Yang-Mills theory is%
\begin{equation}
g_{eff}\sim g_{s}N_{c}u^{p-3},
\end{equation}%
which is $u$ dependent. This $u$ dependence corresponds to the RG flow in
the Yang-Mills theory, i.e. the effective $g_{eff}$\ coupling constant
depends on the energy scale $u$. In the case of D3-brane, $g_{eff}\sim
g_{s}N_{c}$\ becomes a constant and the dual Yang-Mills theory is $\mathcal{N%
}=4$ SYM theory which is a conformal field theory. The curvature\ of the
background (\ref{metric}) is%
\begin{equation}
\mathcal{R}\sim \frac{1}{l_{s}^{2}g_{eff}},
\end{equation}%
which reflects the string/gauge duality - the string on a background of
curvature $\mathcal{R}$\ is dual to a gauge theory with the effective
coupling $g_{eff}$. To make the perturbation valid in the string side, we
require that the curvature is small $\mathcal{R}\ll 1$, which means that the
effective coupling in the dual gauge theory is large $g_{eff}\gg 1/l_{s}^{2}$%
. In the case of D3-brane, the curvature $\mathcal{R}$\ becomes a constant, and 
the background (\ref{metric}) reduces to a constant curvature spacetime - $%
AdS_{5}\times S^{5}$.

The coordinates transformation (for the cases of $p\neq 5$)%
\begin{equation}
u=\left( \frac{5-p}{2}\right) ^{\frac{2}{p-5}}R^{\frac{p-7}{p-5}}z^{\frac{2}{%
p-5}},
\end{equation}%
brings the above solution (\ref{metric}) to the following \emph{Poincar\'{e}
form,}%
\begin{equation}
ds^{2}=e^{2A\left( z\right) }\left[ \eta _{\mu \nu }dx^{\mu }dx^{\nu
}+dz^{2}+\frac{\left( p-5\right) ^{2}}{4}z^{2}d\Omega _{8-p}^{2}\right] .
\end{equation}%
Next, we consider $N_{f}$\ probe $Dq$-branes with $q-4$ of their dimensions
in the $S^{q-4}$ part of $S^{8-p}$, with the other dimensions in $z$ and $%
x^{\mu }$ directions as given in Table \ref{embed}, 
\begin{table}[th]
\begin{center}
$%
\begin{tabular}{|c|c|c|c|c|c|c|c|c|c|}
\hline
& 0 & 1 & 2 & 3 & $\cdots$ & $p$ & $z$ & \multicolumn{2}{|c|}{$S^{8-p}$} \\ 
\hline
D$p$ & $\bullet$ & $\bullet$ & $\bullet$ & $\bullet$ & $\bullet$ & $\bullet$
& - & \multicolumn{2}{|c|}{-} \\ \hline
D$q$ & $\bullet$ & $\bullet$ & $\bullet$ & $\bullet$ & - & - & $\bullet$ & $%
S^{q-4}\subset S^{8-p}$ & - \\ \hline
\end{tabular}
\ $%
\end{center}
\caption{ Spacetime embedding of $D_{p}-D_{q}$ system.}
\label{embed}
\end{table}
The induced $q+1$ dimensions metric on the probe branes is given as 
\begin{equation}
ds^{2}=e^{2A}\left[ \eta _{\mu \nu }dx^{\mu }dx^{\nu }+dz^{2}+\frac{z^{2}}{%
z_{0}^{2}}d\Omega _{q-4}^{2}\right] .  \label{induced}
\end{equation}

Where\ $\mu,\nu=0,\cdots,3$, and the metric function of the warp factor only
includes the logarithmic term 
\begin{align}
A(z)=-a_{0} ~\mathrm{ln} z, ~~with ~~a_{0}=\frac{p-7}{2\left( p-5\right) },
\end{align}
and the dilaton field part takes the form of 
\begin{equation}
e^{\Phi} = \ g_{s}\left( \frac{2}{5-p}\frac{R}{z}\right) ^{\frac{\left(
p-3\right) \left( p-7\right) }{2\left( p-5\right) }}.
\end{equation}
It follows that 
\begin{equation}
\Phi(z) \sim d_{0} \ln z, ~with ~d_{0}=-\frac{\left( p-3\right) \left(
p-7\right) }{2\left( p-5\right) }.  \label{Phi}
\end{equation}

\subsection{The deformed $D_p-D_q$ soft-wall model}

\label{subsec-softmodel}

In the above subsection, we have derived the general metric structure of the 
$D_p-D_q$ system in Type II superstring theory, and we have noticed that the
metric function $A(z)$ only includes the logarithmic term, and in general
there is another logarithmic contribution to the dilaton field. However,
from the lessons of AdS$_5$ metric ($D_3$ system) and the Sakai-Sugimoto
model($D_4-D_8$ system), the $D_p-D_q$ system cannot describe linear
trajectories of mesons. It was shown in Ref.\cite{KKSS2006}, in order to
produce linear trajectories, there should be a $z^2$ term, but all $z^2$
asymptotics should be kept in the dilaton field $\Phi(z)$ and not in the
warp factor $A(z)$. Otherwise, the radial slope $a_n$ will be spin
dependent. Therefore, to describe the real QCD, we propose a deformed $%
D_p-D_q$ soft-wall model which is defined as 
\begin{equation}
A(z)=-a_0 ~\mathrm{ln} z, \, ~\Phi(z)=d_{0}\ln z+c_{2}z^{2}\,.
\label{scenario1}
\end{equation}

By assuming that the gauge fields are independent of the internal space $%
S^{q-4}$, after integrating out $S^{q-4}$, up to the quadratic terms and
following the same assumption as in \cite{KKSS2006}, we can have the
effective 5D action for higher spin mesons described by tensor fields as 
\begin{align}
I & =\frac{1}{2}\int d^{5}x\sqrt{g}\,\,e^{-\Phi(z)}\bigg \{%
\Delta_{N}\phi_{M_{1}\cdots M_{S}}\Delta^{N}\phi^{M_{1}\cdots M_{S}}  \notag
\\
& +m_{5}^{2}\phi_{M_{1}\cdots M_{S}}\phi^{M_{1}\cdots M_{S}}\bigg \},
\label{spin}
\end{align}
where $\phi_{M_{1}\cdots M_{S}}$ is the tensor field and $M_i$ is the tensor
index. The value of $S$ is equal to the spin of the field. The parameter $%
m_{5}^2$ is the 5D mass square of the bulk fields, and $g$ and $\Phi(z)$ are
the induced $q+1$ dimension metric and dilaton field as shown in Eq. (\ref%
{induced}) and (\ref{Phi}). The action for $\rho_1, a_1$ and $\rho_3, a_3$
mesons is given by taking $S=1$, and $S=3$ respectively.

Following the standard procedure of dimensional reduction, we can decompose
the bulk field into its 4d components and their fifth profiles as $%
\phi(x;z)_{M_{1}\cdots M_{S}}=\sum_{n=0}\phi^n_{M_{1}\cdots M_{S}}(x)
\psi_n(z)$. The equation of motion (EOM) of the fifth profile wavefunctions $%
\psi(z)_n$ for the general higher spin field can be derived as 
\begin{eqnarray}
\partial_z^{2}\psi_{n} - \partial_z B \cdot \partial_z\psi_{n} +\left(
M_{n,S}^{2} -m_{5}^{2}e^{2A}\right) \psi_{n} =0\,,  \label{hispin}
\end{eqnarray}
where $M_{n,S}$ is the mass of the 4-dimension field $\phi^{n}_{M_{1}\cdots M_{S}}(x)$ and%
\begin{eqnarray}
B=\Phi-k (2\,S-1) A= \Phi+c_0 (2\,S-1) \mathrm{ln}z
\end{eqnarray}
is the linear combination of the metric background function and the dilaton
field. The combination function $B(z)$ approaches logarithmic asymptotic at UV
brane, and goes to $z^2$ asymptotic at IR region.
It is worthy of remark that the spin parameter $S$ enters in the
factor $B$ and can affect the EOM and spectra. Two comments are in order: 1)
The EOM for the eigenspectrum and wavefunction is valid for all mesons. 2)
The spin parameter $S$ enters in the factor $B$ and can affect the EOM and
spectra.

The parameter $k$ is a parameter depending on the induced metric (\ref%
{induced}) of the $D_q$ brane. After integrating out $S^{q-4}$, $k$ is
determined as 
\begin{equation}
k=-\frac{\left( p-3\right) \left( q-5\right) +4}{7-p}.
\end{equation}
It is obviously that $k$ depends on both $p$ and $q$. For simplicity, we
have defined 
\begin{equation}
c_0=k a_0= -\frac{\left( p-3\right) \left( q-5\right) +4}{2\left( p-5\right) 
}.
\end{equation}

The parameters $a_{0},d_{0}$, the other two parameters $k,c_{0}$ and
corresponding curvature $\mathcal{R}$ for any $Dp-Dq$ system are listed in
Table \ref{c0d0}. 
\begin{table}[th]
\begin{center}
$%
\begin{tabular}{|c|c|c|c|c|c|c|c|}
\hline
$p$ & \multicolumn{2}{|c|}{$3$} & \multicolumn{3}{|c|}{$4$} & 
\multicolumn{2}{|c|}{$6$} \\ \hline
$q$ & \multicolumn{1}{|c|}{$5$} & $7$ & $4$ & $6$ & $8$ & $4$ & $6$ \\ \hline
$k=-\frac{\left( p-3\right) \left( q-5\right) +4}{p-7}$ & 
\multicolumn{2}{|c|}{$1$} & $1$ & $5/3$ & $7/3$ & $-1$ & $-7$ \\ \hline
$a_{0}=\frac{p-7}{2\left( p-5\right) }$ & \multicolumn{2}{|c|}{$1$} & 
\multicolumn{3}{|c|}{$3/2$} & \multicolumn{2}{|c|}{$-1/2$} \\ \hline
$c_{0}=ka_{0}$ & \multicolumn{2}{|c|}{$1$} & $3/2$ & $5/2$ & $7/2$ & $1/2$ & 
$7/2$ \\ \hline
$d_{0}=-\frac{\left( p-3\right) \left( p-7\right) }{2\left( p-5\right) }$ & 
\multicolumn{2}{|c|}{$0$} & \multicolumn{3}{|c|}{$-3/2$} & 
\multicolumn{2}{|c|}{$3/2$} \\ \hline
$\mathcal{R}\sim \frac{1}{g_{eff}}$ & \multicolumn{2}{|c}{$1/\sqrt{3}$} & 
\multicolumn{3}{|c}{$z^{-2}/\sqrt{36\pi }$} & \multicolumn{2}{|c|}{$6\sqrt{2}%
z^{6}$} \\ \hline
\end{tabular}%
\ $%
\end{center}
\caption{Theoretical results for the $D_{p}-D_{q}$ system.}
\label{c0d0}
\end{table}
We notice that $d_{0}=0$ for $D_{3}$ background branes, \textit{i.e.}
dilaton field is constant in AdS$_{5}$ space. However, the dilaton field in
a general $Dp-Dq$ system can have a $\ln z$ term contribution, e.g. in the $%
D4-D8$ system (Sakai-Sugimoto model \cite{SS}), $d_{0}=-3/2$.
We also want to point out that for pure $Dp-Dq$ system, the curvature is
proportional to the inverse of the coupling strength $g_{eff}$. For $D_{3}$
background branes, the curvature is a constant. The curvature for $D_{4}$
background branes is small at IR, and large at UV,  its dual gauge theory 
is strongly coupled at IR and weakly coupled at UV, which is similar to QCD.
However, the curvature for $D_{6}$
background branes is large at IR, and small at UV,  its dual gauge theory 
is weakly coulped at IR and strongly coupled at UV, which is opposite to QCD.

\section{Determine \textit{Holographic} QCD model from Regge trajectories of 
$\protect\rho_1$}

\label{sec-rho-metric}

In this section we firstly consider the spectra of vector mesons, and we
will investigate axial-vectors mesons and the chiral symmetry breaking
mechanism in more detail in Sec. \ref{sec-a-metric}.

In the dictionary of AdS/CFT, a $f-$form operator with conformal dimension $%
\Delta$ has 5-dimensional square-mass $m_5^2=(\Delta -f)(\Delta + f -4)$ in
the bulk \cite{5Dmass}, and for vector mesons $m_{5}^{2}=0$. The spectra of
EOM of Eq. (\ref{hispin}) for vector mesons has an exact solution: 
\begin{equation}
M_{n,S}^{2}=4c_{2}n+4c_{2}c_{0}S+2c_{2}(1-c_{0}+d_{0})\,.  \label{exact11}
\end{equation}
When $c_{0}=c_{2}=1$ and $d_{0}=0$, this solution reduces to the results
given in the Ref. \cite{KKSS2006}.

This exact solution supports the parameterization on Regge trajectories and
can tell how phenomenological parameters $a_{n}$, $a_{S}$ and $b$ are
directly related with the metric parameters $c_{0}$, $c_{2}$ and $d_{0}$:

1) $c_{2}$ is completely determined by the string tension in the radial
direction $a_n$, \textit{i.e.} 
\begin{equation}
c_{2}=\frac{a_{n}}{4}.
\end{equation}
Andreev in Ref. \cite{Andreev-z2} shows that there is an upper bound of $%
c_{2}$. $c_{2}$ can be determined by the coefficient $C_{2}$ \cite{c2} of
the quadratic correction to the vector-vector current correlator \cite{svz} 
\begin{equation}  \label{PiVq2}
\mathcal{N} q^{2}\,\frac{d \Pi_{V} }{dq^{2}}=C_{0}+\frac{1}{q^{2}}C_{2}+
\sum_{n\geq2}\frac{n}{q^{2n}}C_{2n}\langle\mathcal{O}_{2n}\rangle\,,
\end{equation}
where $C_{2}$ can be determined from $e^{+} e^{-}$ scattering data \cite%
{Narison:1992ru}. According to \cite{Andreev-z2}, the relation between $%
C_{2} $ in Eq. (\ref{PiVq2}) and the parameter $c_{2}$ for Dilaton field is: 
$c_{2} = - \frac{3}{2} C_{2}$. The experimental bound $|\,C_{2}\,|\leq0.14 \,%
\text{GeV}^{\,2}$ gives $|\,c_{2}\,|\leq0.21 \,\text{GeV}^{\,2}$. From the
fitting result for the Regge trajectories of vector and axial-vector mesons
Eqs. (\ref{eq:rhoregge}) and (\ref{eq:aregge}), we can read that $c_2$ is
around $0.2\sim 0.25$, which is in agreement the experimental upper bounds.

2) It is interesting to notice that $c_{0}$ reflects the difference of
string tension in the radial direction and spin direction, 
\begin{equation}
c_{0}=\frac{a_{S}}{a_{n}}.
\end{equation}
From the string theory side, it is commonly believed that the dual string theory of 
describing QCD should be strongly curved at high energy scales and weakly curved 
at low energy scales \cite{Aharony-curvature}. It seems $D4$ background brane system
is more like the dual string theory of QCD. However, by reading
the result for the Regge trajectories of vector and axial-vector mesons Eqs.
(\ref{eq:rhoregge}) and (\ref{eq:aregge}), we can see that $a_S/a_n$ is
around $1$, which indicates that the real \textit{holographic} QCD model is
not far away from $AdS_5$ model.

3) Furthermore, $d_0$ can be solved out as 
\begin{equation}
d_0=\frac{2 b}{a_n} + \frac{a_S}{a_n} - 1.
\end{equation}
If we take the approximation of $a_n=a_S=1$, we have $c_0=1, c_2=1/4$ for
both vector and axial-vector mesons, while $d_0$ is mainly determined by the
ground state square-mass as $d_0^{\rho/a}= 2b^{\rho/a}$.

From the above analysis, we can see that the realistic \textit{holographic}
QCD model should be close to models defined in D$p$-branes background for $%
p=3$, i.e., $AdS_5$ background. By using Eq. (\ref{exact11}) to fit the
spectra of vector mesons $\rho_1$, we get the following metric parameters: 
\begin{equation}  \label{metric-rho}
\begin{array}{rl}
c_{0}^{\rho}=\!\!\! & 1 , \\ 
c_{2}^{\rho}=\!\!\! & 0.2 \mathrm{GeV}^{2}, \\ 
d_{0}^{\rho}=\!\!\! & 0 .%
\end{array}%
\end{equation}
It is noticed that here, $c_2$ has dimension $\mathrm{GeV}^2$. In doing
numerical simulation, we have introduced a dimensionless parameter $u$. The
relation between $u$ and $z$ is defined as 
\begin{align}
u = \Lambda_{Scl} \,\, z \,\,.
\end{align}
In our fitting, we have fixed $\Lambda_{Scl}=0.2$, $u_{UV}=0.1$ and $%
u_{UV}=3 $, respectively. We will use the same parameters for calculations
of axial-vector mesons.

\section{Axial vector mesons in the deformed $AdS_5$ model}

\label{sec-a-metric}

We have determined the metric structure Eq. (\ref{metric-rho}) of describing
the vector mesons, and we suppose the spectra of axial-vector mesons $a_1$
can be described in the same \textit{holographic} QCD model. As mentioned in
the introduction, in order to describe the Regge trajectories of
axial-vector mesons $a$, we have to describe the chiral symmetry breaking in
the vacuum and the asymptotic chiral symmetry restoration in highly excited
meson states.

\subsection{Chiral symmetry breaking}

In order to produce the splitting of axial-vector meson spectra from vector
meson spectra, it is essential to know the chiral symmetry breaking
mechanism, in the following we discuss two different ways of chiral symmetry
breaking.

\subsubsection{Constant $m_{5,a}^2$ in the Higgsless model}

\label{sec-a1-m5}

If we start from the Higgsless model Eq. (\ref{spin}), the only difference
between the EOM for $\rho$ and $a$ is the 5D bulk mass $m_5^2$. We have
taken $m_{5}^2=0$ for vector mesons, we can assume that $m_{5}^2 \neq 0$ for
axial-vector mesons due to chiral symmetry breaking.

For the general $A(z)$ and $\Phi(z)$ parameterized as 
\begin{eqnarray}
A(z) &=& - c_0 \ln z \,, \\
\Phi(z) &=& c_2 z^2\,,
\end{eqnarray}
the potential derived from $B_s(z) = \Phi (z) - (2 S - 1) A(z) $ can be put
as 
\begin{eqnarray}
V(z) = 2 c_2 (m -1) + \frac{m^2 - \frac{1}{4} }{z^2} + c_2^2 z^2 + \frac{%
m_5^2}{ z^{2 c_0} }\,,  \label{V-m5}
\end{eqnarray}
with 
\begin{eqnarray}
m = [(2 S - 1 ) c_0 + 1]/2 \,.
\end{eqnarray}

When $c_0=1$, for any $c_2$ and constant $m_5$, the EOM 
\begin{eqnarray}
- \Psi^{\prime \prime} + V(z) \Psi = m_n^2 \Psi \,.  \label{EOM-a1}
\end{eqnarray}
can have analytic solution: 
\begin{eqnarray}
m_n^2 = c_2 (4 n + 1) + c_2 ( 2 S -1) + c_2 \sqrt{[(2 S -1) + 1]^2 + 4 m_5^2 
}\,.
\end{eqnarray}
When $S=1$, 
\begin{eqnarray}
m_n^2 =4 c_2 n + 4 c_2 + 2 c_2 \left(\sqrt{1 + m_5^2 }-1 \right) \,.
\label{spectra-m5}
\end{eqnarray}

\begin{figure}[t]
\centerline{
\epsfxsize=8.0 cm \epsfysize=7.5cm \epsfbox{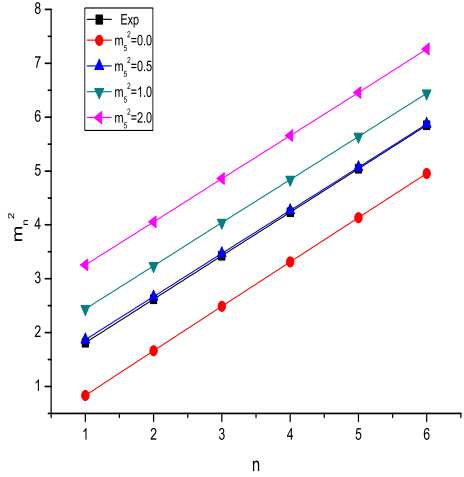}
}
\caption{\textit{The spectra of axial-vector mesons $a_1$ by solving the EOM
of $a_1$ Eq. (\protect\ref{EOM-a1}) for different $m_{5}^2$.}}
\label{fig-a1-m5}
\end{figure}

From Eq. (\ref{spectra-m5}), for $m_{5,\rho}^2=0$ and $m_{5,a}^2 \neq 0$, we
can get the spectra of axial-vector $a_1$ through shifting the spectra of $\rho_1$
by $2 c_2 (\sqrt{1 + m_{5,a}^2 }-1)$ upward. In Fig. \ref{fig-a1-m5}, we
explicitly show our numerical results of the spectra of axial-vector mesons $%
a_1$ by solving the EOM of $a_1$ Eq. (\ref{EOM-a1}) for different $m_{5}^2$.
It is found that when $m_{5,a}^2=0.5 \mathrm{GeV}^2$, the produced $a_1$
spectra agrees well with the experimental data.

\subsubsection{Higgs Mechanism}

In Ref. \cite{KKSS2006}, it is suggested that the axial field picks up a $z$%
-dependent 5d mass via the Higgs mechanism from the background scalar $X$
that encodes the chiral symmetry breaking. The axial vector meson mode
equation reads: 
\begin{equation}  \label{axialmode}
\partial_z \left ( e^{-\Phi(z)} e^{A(z)} \partial_z a_n \right ) + [m_n^2 -
g_5^2\, e^{2A(z)}\, X(z)^2] e^{-\Phi(z)} e^{A(z)} a_n(z) = 0\, ,
\end{equation}
with $g_5^2=12 \pi^2/N_c$. The linearized equation of motion for the scalar
field $X$ reads: 
\begin{equation}  \label{vev}
\partial_z \left ( e^{-\Phi(z)} e^{3 A(z)} \partial_z X(z) \right ) + 3
e^{-\Phi(z)} e^{5 A(z)} X(z) = 0\, .
\end{equation}
It is quite complicated to solve the coupled Eqs. (\ref{axialmode}) and (\ref%
{vev}). We only discuss two asymptotic solutions of scalar field at UV ($z=0$%
) and IR, respectively.

\begin{figure}[t]
\centerline{
\epsfxsize=6.5 cm \epsfysize=6.0 cm \epsfbox{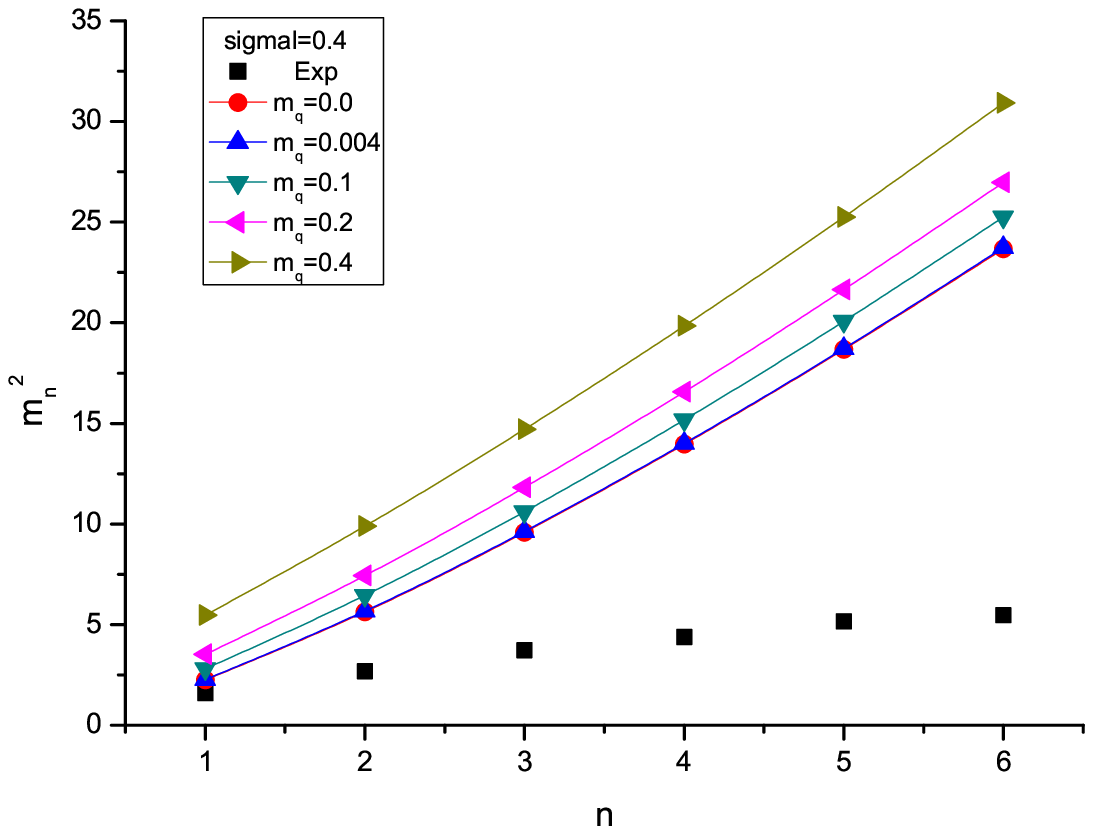}
\hspace*{0.1cm}
\epsfxsize=6.5 cm \epsfysize=6.0cm \epsfbox{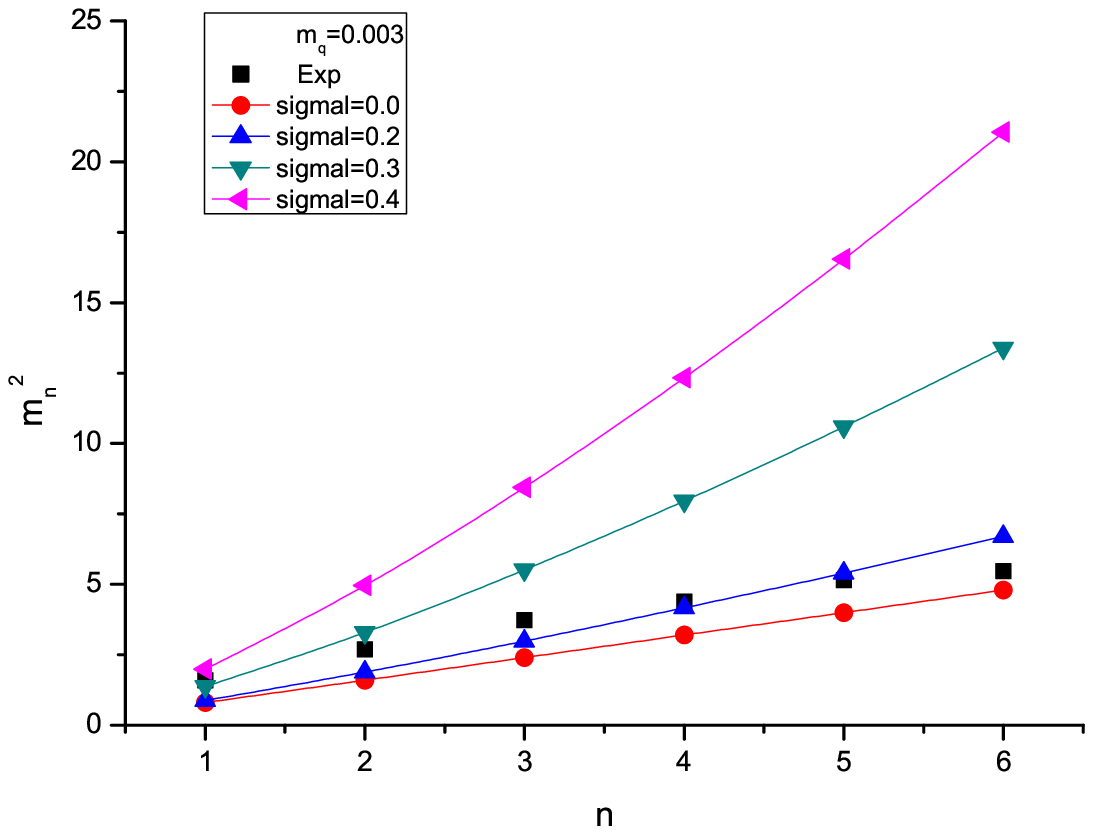}
} \vskip -0.05cm \hskip 0.4 cm \textbf{( a ) } \hskip 5.8cm \textbf{( b )}
\caption{\textit{The $a_1$ spectra by using the UV asymptotic form of scalar
field solution at $N_c=3$, (a) is for different values of $m_q$ with fixed $%
\Sigma=(400 \mathrm{MeV})^3$, (b) is for different values of $\Sigma$ with
fixed $m_q=3 \mathrm{MeV})$.}}
\label{fig-a1-higgs}
\end{figure}

The asymptotic form at UV has the form of 
\begin{equation}  \label{XUV}
X(z)\ \overset{z\to0}{\to}\ \frac{1}{2} m_q z + \frac{1}{2}\Sigma z^3,
\end{equation}
here the coefficient $m_q$ is the UV ($z=0$) boundary condition given by the
quark mass matrix, while the coefficient $\Sigma$ is the chiral condensate,
which can be determined dynamically by the boundary condition in the IR. We
show the produced $a_1$ spectra from Eq. (\ref{XUV}) in Fig. \ref%
{fig-a1-higgs} when $N_c=3$. Fig. \ref{fig-a1-higgs}(a) is for different
values of $m_q$ with fixed $\Sigma=(400 \mathrm{MeV})^3$, and Fig. \ref%
{fig-a1-higgs}(b) is for different values of $\Sigma$ with fixed $m_q=3 
\mathrm{MeV}$. It can be clearly seen that the asymptotic form at UV
introduces nonlinearity of the $n$ dependence of $M_{n}^2$, the $z^3$ term
contributes more on the non-linear behavior.

For large $z$ (in the IR) on the background $\Phi=c_2 z^2$ the equation for $%
X$ becomes 
\begin{equation}
X^{\prime \prime }- 2 c_2 z X^{\prime }+ \frac3{z^2}X = 0 \qquad(z\gg1),
\end{equation}
the scalar field has a solution goes to a constant in the IR, i.e., $%
X(z)\rightarrow const$ as $z\rightarrow \infty$. $X(z)= const$ means the
effective 5D mass $m_{5,a}^2= g_5^2\, X(z)^2$ is a constant. Then the
spectra of $a_1$ will behave the same as that in Sec. \ref{sec-a1-m5}.

\subsection{Chiral Symmetry restoration}

Though the slopes in radial direction for vector mesons $a_n^{\rho}$ and
axial-vector mesons $a_n^{a}$ can be roughly taken as the same, i.e, $%
a_n^{\rho}\simeq a_n^{a}$, in order to have chiral symmetry restoration, we
need $a_n^\rho > a_n^{a}$. To accommodate such a case, we can take the
ansatz that the effective 5D mass $m_5^2(z)$ has the following form 
\begin{eqnarray}
\frac{ m_5^2}{z^2} = \delta c_2^2 z^2 + \frac{\delta S^2}{z^2} + \delta
m_0^2\,.
\end{eqnarray}
Substituting this ansatz into Eq. (\ref{V-m5}), we can find the spectra can
be exactly solved out as 
\begin{eqnarray}
m_n^2 = 4 c_2^\prime n + 2 c_2 S + 2 c_2^\prime S^\prime + \delta m_0^2 + 2
(c_2 + c_2^\prime)\,,
\end{eqnarray}
where $c_2^\prime = \sqrt{c_2^2 + \delta c_2^2}$ and $S^\prime=\sqrt{S^2 +
\delta S^2}$. Then equivalently, the chiral symmetry restoration demands
that $c_2^\prime$ should be smaller than $c_2$, which infers that $\delta
c_2^2$ should be negative. It is not clear at the moment, how to get the negative
correction $\delta c_2^2 z^4$ in 5D mass square of axial-vector meson from 
Higgs mechanism, which deserves further careful study in the future.

\section{Summary}

\label{sec-summary}

In summary,  we have derived the general 5-dimension metric structure of the 
$Dp-Dq$ system in type II superstring theory. We have shown the dependence 
of the metric parameters on the Regge trajectories parameters: The quadratic 
term in the dilaton background field is solely determined by the slope in the 
radial direction; The warp factor is mainly determined by the difference of the 
slope in the spin direction and the radial direction; The logarithmic term in the
dilaton background field contributes to  the ground state square-masses.
 
It is commonly believed that the dual string theory of describing QCD should be 
strongly curved at high energy scales and weakly curved  at low energy scales,
it seems that the $D4$ background brane system is more like the dual string theory
of QCD. However, the ratio of the slope in the spin direction of the vector mesons 
over its slope in the radial direction $a_S/a_n$ is around $1$, which indicates that 
the real \textit{holographic} QCD model is not far away from $AdS_5$ model.

We  have shown that the spectra of axial vector mesons can also be described in the 
$AdS_5$ soft-wall model, and a constant 5D bulk mass for axial-vector meson 
plays efficient role to realize the chiral symmetry breaking in the vacuum, and a 
small negative $z^4$ correction in the 5D mass square is  helpful to realize the 
chiral symmetry restoration in high excitation states.

The information in this study is important for a realistic holographic QCD model and our
future study on the interactions of mesons (branching ratios and decay
widths of mesons, their interactions, etc). In our current approach, we
chose the Higgsless model to describe the radial and higher spin excitations
of vector and axial-vector mesons, the pseudoscalar $\pi$ is the zero mode
of axial-vector field and there is no its radial and higher spin
excitations. We leave the study of the Regge trajectories of scalar and
pseudo-scalar mesons by using the global symmetry breaking model in our
future works.

\vskip3mm \textit{\textbf{Acknowledgments: ---}} We thank F.L. Lin, J. Liu,
H.C. Ren and J. Shock for discussions. The work of M.H. is supported by 
CAS program "Outstanding young scientists abroad brought-in", the CAS key
project under grant No. KJCX3-SYW-N2, and NSFC under grant No. 10875134 and
No. 10735040. The work of Q.Y. is supported by the NCS of Taiwan (No. NSC
95-2112-M-007-001 and 96-2628-M-007-002-MY3). The work of Y.Y. is supported
by National Science Council (NCS) of Taiwan (97-2112-M-009-019-MY3) and National
Center for Theoretical Sciences (NCTS) through NCS of Taiwan.

\end{document}